\documentclass[aps,pra,showpacs,floatfix,twocolumn]{revtex4}
\usepackage{graphicx} 
\usepackage{amsmath}
\usepackage{bm}
\usepackage{dcolumn}
\usepackage{longtable}

\begin{document}
\newcolumntype{m}[1]{D{.}{.}{#1}}

\title{Second-order effects on the hyperfine structure of $P$ states of alkali-metal atoms}

\author{K. Beloy and A. Derevianko}
 \affiliation{Physics Department, University of
Nevada, Reno, Nevada  89557}

\date{\today}
\begin{abstract}
We analyze second-order $M1$-$M1$ and $M1$-$E2$ effects to the
hyperfine structure (HFS) of the lowest energy $P$ states of
alkali-metal atoms arising from the coupling of the two
($J=1/2,3/2$) fine-structure levels through the hyperfine
interaction. We find these effects to be especially sizable in Li,
leading to a $9\sigma$ correction to the most accurate reported
experimental value of the $A(P_{1/2})$ HFS constant of $^7$Li [D.
Das and V. Natarajan, J. Phys. B \textbf{41}, 035001 (2008)]. For
the remaining alkali-metal systems, the results tabulated within may
be referenced as higher precision is sought in experimental
determination of the HFS constants.
\end{abstract}

\maketitle

\section{Introduction}
As experimental accuracy improves, the interpretation of
measurements may require refined theoretical analysis. Measurements
of the hyperfine structure (HFS) is one of such examples. The HFS of
atomic systems arises from the coupling between the atomic electrons
and the nuclear spin. Generally speaking, in the approximation that
$J$, representing the electronic angular momentum, remains a
``good'' quantum number, the HFS can be accurately described by the
conventional (first-order) HFS constants $A, B, C,
\dots$~\cite{Arm71}. Respectively, these constants describe the
electronic interaction with the nuclear magnetic dipole ($M1$),
electric quadrupole ($E2$), magnetic octupole ($M3$),\dots~; these
interactions are collectively referred to as the hyperfine
interaction (HFI). Likewise, with $J$ being a ``good'' quantum
number, experimental measurements of the HFS intervals, along with
known values of the nuclear moments, can be used to definitively
determine these HFS constants. However, for atomic states that are
part of a fine-structure manifold, the nearby fine-structure levels
may provide sizable contamination to the
$J$-purity~\cite{AriIngVio77,OrtAckOtt75,BelDerJoh08,BelDerDzu08}.
In these situations, to make the connection between HFS interval
measurements and HFS constants, it becomes necessary to consider the
HFI coupling between fine-structure states~\cite{Arm71}. In terms of
perturbation theory, this arises in the second-order in the HFI and
is dominated by the $M1$-$M1$ and $M1$-$E2$ effects.

In this paper, we provide results for calculations of the
second-order corrections to the HFS for the lowest energy $P$ states
of the naturally occurring alkali-metal isotopes. These isotopes,
along with their respective nuclear properties, are displayed in
Table~\ref{Tbl:nucprops}. The HFS constants $A(P_{1/2})$,
$A(P_{3/2})$, and $B(P_{3/2})$ have been experimentally measured for
all of these isotopes (see
Refs.~\cite{AriIngVio77,DasNat08,FalTieLis06} and references
within).
In a recent paper, Das and Natarajan~\cite{DasNat08} claim a
significant improvement in experimental uncertainty over earlier
works for most of these constants.

\begin{table}[h]
\caption{Nuclear properties of stable alkali-metal isotopes. $I^\pi$
represents the nuclear spin and parity. Nuclear dipole and
quadrupole values are given in terms of nuclear magnetons ($\mu_N$)
and barns (b), respectively. All data is taken from
Ref.~\cite{Sto05}.}\label{Tbl:nucprops}
\begin{ruledtabular}\begin{tabular}{ccm{1.12}m{2.8}}
Isotope & $I^\pi$ & \multicolumn{1}{c}{$\mu~(\mu_N)$}&
\multicolumn{1}{c}{$Q~(\text{b})$} \\
\hline
$^{  6}$Li & $  1^+$ & 0.8220473(6)   & -0.00082(2) \\
$^{  7}$Li & $3/2^-$ & 3.2564625(4)   & -0.0406(8)  \\
$^{ 23}$Na & $3/2^+$ & 2.2176556(6)   &  0.1045(10) \\
$^{ 39}$K  & $3/2^+$ & 0.39150731(12) &  0.0585(6)  \\
$^{ 41}$K  & $3/2^+$ & 0.21489274(12) &  0.0711(7)  \\
$^{ 85}$Rb & $5/2^-$ & 1.3533515(8)   &  0.277(1)   \\
$^{ 87}$Rb & $3/2^-$ & 2.751818(2)    &  0.134(1)   \\
$^{133}$Cs & $7/2^+$ & 2.5829128(15)  & -0.00355(4) \\
\end{tabular}\end{ruledtabular}
\end{table}

Due to its relatively small fine-structure splitting, the
second-order effects of the HFI are most noticeable in Li, the
lightest of the alkali-metal atoms. Over thirty years ago, in
experimental determination of the HFS constants, Orth {\it et
al.}~\cite{OrtAckOtt75,AriIngVio77} recognized the need to consider
the mixing of the fine-structure $P$ states of Li. However, more
recent measurements of the $^{6,7}$Li $A(P_{1/2})$
constants~\cite{WalAshCla03,DasNat08} have overlooked this important
effect. Namely, we find that the second-order correction to the
$^{7}$Li $A(P_{1/2})$ constant causes a shift which is a full order
of magnitude more than claimed uncertainty of Das and
Natarajan~\cite{DasNat08}. For the remaining alkali-metal isotopes,
we find that the second-order effects are not significant at the
current levels of experimental precision for the HFS constants.
However as higher precision is sought, it may become necessary to
include these effects. The compilation of second-order corrections
provided herein may provide useful complementary data for future
high-precision experiments on these isotopes.


This paper is organized as follows. In Section~\ref{Sec:HFI-PT}, we
present pertinent equations for the HFI and corresponding first- and
second-order energies obtained from a standard perturbation theory
analysis. In Section~\ref{Sec:HFI-tens} we provide a reformulated
tensorial analysis of the perturbation theory; this formalism
provides more insight into the rotational symmetries of the various
second-order contributions. We compile HFS expressions for the $P$
states of the alkali-metal atoms in Section~\ref{Sec:HFSeqns}. In
Section~\ref{Sec:Results} we tabulate our results, followed by
concluding remarks in Section~\ref{Sec:Disc}. Furthermore, we
include an Appendix containing notations and expressions for
spherical tensor operators appearing in Sections~\ref{Sec:HFI-PT}
and~\ref{Sec:HFI-tens}.

\section{The hyperfine interaction}\label{Sec:HFI-PT}
In this Section, we recapitulate the basic properties of the
hyperfine interaction and the application of perturbation theory to
analyze its effects on atomic structure. This Section is an
abbreviated review of the presentation provided in
Ref.~\cite{BelDerJoh08}. We employ the notation
$\langle{T_{k}}\rangle_{I}$ to represent the expectation value of
the zero-component operator of spherical tensor (of rank $k$)
${T_{k}}$ in the ``stretched'' state $|I,M_{I}=I\rangle$. This is
related to the reduced matrix element through the expression
\begin{equation*}
\langle{T_{k}}\rangle_{I} =\left(
\begin{array}
[c]{ccc}%
I & k & I\\
-I & 0 & I
\end{array}
\right)\langle I||{T_{k}}||I\rangle.
\end{equation*}

The hyperfine interaction can be expressed as a sum over scalar
products of spherical tensors:
\begin{equation}
\label{Eq:HFI_sp}%
H_{\text{HFI}}=\sum_{k}T_{k}^{n}\cdot T_{k}^{e},
\end{equation}
where $T_{k}^{n}$ and $T_{k}^{e}$ are spherical tensors of rank $k$
($k>0$) acting in the nuclear and electronic spaces, respectively.
As $H_{\text{HFI}}$ is a scalar operator in the combined space, it
is convenient to work in the conventional basis formed by coupling
nuclear, $\left\vert IM_{I}\right\rangle$, and electronic,
$\left\vert \gamma JM_{J}\right\rangle$, states,
\begin{equation}
|\gamma
IJFM_{F}\rangle=\sum_{M_{J},M_{I}}C_{IM_{I};JM_{J}}^{FM_{F}}\left\vert
IM_{I}\right\rangle\left\vert \gamma JM_{J}\right\rangle  ,
\label{Eq:basis}
\end{equation}
with $\gamma$ encapsulating remaining electronic quantum numbers and
the coupling coefficients being the Clebsch-Gordon coefficients. A
matrix element of the HFI between these basis states is given by
\begin{widetext}
\begin{equation*}
\langle\gamma^{\prime}IJ^{\prime}F^{\prime}M_{F}^{\prime}|H_{HFI}|\gamma
IJFM_{F}\rangle =\delta_{F^{\prime}F}\delta_{M_{F}^{\prime}M_{F}}
(-1)^{I+J^\prime+F} \sum_{k}\left\{
\begin{array}
[c]{ccc}%
 I          & J & F \\
 J^{\prime} & I & k
\end{array}
\right\}  \,
\langle I||T_{k}^{n}||I\rangle\langle\gamma^{\prime}J^{\prime}||T_{k}^{e}||\gamma J%
\rangle.\label{Eq:HFIgen}
\end{equation*}

First-order and second-order energy corrections to the state
described by electronic quantum numbers $\gamma$ and $J$ are
\begin{eqnarray}
W^{(1)}_{F}&=&\langle{\gamma}IJFM_{F}|H_{HFI}|\gamma IJFM_{F}\rangle
=(-1)^{I+J+F} \sum_{k}\left\{
\begin{array}
[c]{ccc}%
 I & J & F \\
 J & I & k
\end{array}
\right\}  \,
\langle I||T_{k}^{n}||I\rangle\langle\gamma J||T_{k}^{e}||\gamma J%
\rangle,\label{Eq:W1}\\
W^{(2)}_{F}&=&\sum_{\gamma^{\prime}J^{\prime}} \frac{\langle\gamma
IJFM_{F}|H_{HFI}|\gamma^{\prime}IJ^{\prime}FM_{F} \rangle
\langle\gamma^{\prime}IJ^{\prime}FM_{F}|H_{HFI}|\gamma
IJFM_{F}\rangle}{E_{\gamma
J}-E_{\gamma^{\prime}J^{\prime}}}\nonumber\\
&=&\sum_{\gamma^{\prime}J^{\prime}}
\frac{\left(-1\right)^{J-J^\prime}} {E_{\gamma
J}-E_{\gamma^{\prime}J^{\prime}}} \sum_{k_1,k_2} \left\{
\begin{array}{ccc}
 I          & J & F \\
 J^{\prime} & I & k_1
\end{array}
\right\} \left\{
\begin{array}{ccc}
 I          & J & F \\
 J^{\prime} & I & k_2
\end{array}
\right\} \langle I||T_{k_1}^{n}||I\rangle \langle
I||T_{k_2}^{n}||I\rangle \langle\gamma
J||T_{k_1}^{e}||\gamma^{\prime}J^{\prime}\rangle
\langle\gamma^{\prime}J^{\prime}||T_{k_2}^{e}||\gamma J\rangle.
\label{Eq:W2_a}
\end{eqnarray}
\end{widetext}
Here the summations exclude the case
$\left(\gamma^{\prime}J^{\prime}\right)=\left(\gamma J\right)$; this
will be implicit in similar summations to follow. We note here that,
due to large energies associated with nuclear excitations, we treat
the nuclear states as ``good'' quantum states. The reduced matrix
elements $\langle I||T_{k}^{n}||I\rangle$ appearing here are
associated with the nuclear moments. Specifically, the magnetic
dipole ($k=1$), electric quadrupole ($k=2$), and magnetic octupole
($k=3$) are given by
\begin{equation*}
\begin{array}{ccr}
\mu &=&\langle{T_{1}^{n}}\rangle_{I},\\
Q   &=&2\langle{T_{2}^{n}}\rangle_{I},\\
\Omega &=&-\langle{T_{3}^{n}}\rangle_{I}.
\end{array}
\end{equation*}
Furthermore the product of reduced matrix elements $\langle
I||T_{k}^{n}||I\rangle\langle\gamma J||T_{k}^{e}||\gamma J\rangle$
appearing in Eq.~(\ref{Eq:W1}) correspond to the conventional
first-order hyperfine constants. Specifically, through the $C$
constant, these are given by
\begin{equation}
\label{Eq:HFconsts}
\begin{array}{rrrrr}
A &=& \frac{1}{IJ}
      \langle T_{1}^{n}\rangle_{I}\langle T_{1}^{e}\rangle_{J}
  &=&\frac{1}{IJ}\mu\langle T_{1}^{e}\rangle_{J},\\
B &=& 4\langle T_{2}^{n}\rangle_{I}\langle T_{2}^{e}\rangle_{J}
  &=&2 Q\langle T_{2}^{e}\rangle_{J},\\
C &=&\langle T_{3}^{n}\rangle_{I}\langle T_{3}^{e}\rangle_{J}
  &=&-\Omega\langle T_{3}^{e}\rangle_{J}.
\end{array}
\end{equation}
For electronic states within a fine-structure manifold, the leading
second-order effects are due to mixing with the nearby states within
the manifold (i.e., $\gamma^\prime=\gamma$). Furthermore, these
effects are dominated by the dipole-dipole ($M1$-$M1$) and
dipole-quadrupole ($M1$-$E2$) terms. The constants $\eta$ and
$\zeta$ have been used in Refs.~\cite{BelDerJoh08,BelDerDzu08} to
parameterize these effects and are given by
\begin{eqnarray}
\eta&=&\mp\frac{(I+1)(2I+1)}{I}\frac{\mu^2\left\vert\langle\gamma
J||T_{1}^{e}||\gamma{J\pm1}\rangle\right\vert^2}
{E_{\gamma J}-E_{\gamma{J\pm1}}},\label{Eq:eta}\\
\zeta&=&\mp\frac{(I+1)(2I+1)}{I}\sqrt{\frac{2I+3}{2I-1}}
\nonumber\\&&\times\frac{\mu{Q}\langle\gamma
J||T_{1}^{e}||\gamma{J\pm1}\rangle\langle\gamma
J||T_{2}^{e}||\gamma{J\pm1}\rangle}{E_{\gamma
J}-E_{\gamma{J\pm1}}}.\label{Eq:zeta}
\end{eqnarray}
(If both $J\pm1$ levels exist, it would be necessary to distinguish
two independent $\eta$'s and $\zeta$'s.) Explicit formulas for
matrix elements of the electronic tensors $T_{k}^{e}$ are given in
Ref.~\cite{BelDerJoh08}.

\section{Tensorial analysis of second-order effects}\label{Sec:HFI-tens}
In this section we reformulate the second-order contributions to the
energy. We classify the second-order terms according to the
underlying rotational symmetry in the nuclear and electronic spaces.
As a result, specific second-order terms are connected to the
first-order constants $A,B,C,\dots$ It becomes clear from a physical
standpoint why, for instance, the second-order constant $\eta$
cannot affect experimental determination of the $C$ constant, as
proven by brute manner in Ref.~\cite{BelDerJoh08}. However, the
formalism here is also farther reaching, as it can easily be
extended to other second-order terms past $\eta$ and to higher
orders.

We begin with an effective Hamiltonian,
$H^{\text{eff}}_{{\gamma}J}$, which produces the exact hyperfine
energies when acting on the (unperturbed) coupled basis states,
Eq.~(\ref{Eq:basis}), i.e.~$H^{\text{eff}}_{{\gamma}J}|\gamma
IJFM_{F}\rangle=W_F|\gamma IJFM_{F}\rangle$. We decompose this
effective Hamiltonian into contributions of increasing orders of the
HFI,
$H^{\text{eff}}_{{\gamma}J}=H^{(1)}_{{\gamma}J}+H^{(2)}_{{\gamma}J}+\dots$,
with associated energy contributions given by
$W_F^{(m)}=\langle\gamma IJFM_{F}|H^{(m)}_{{\gamma}J}|\gamma
IJFM_{F}\rangle$. From Eq.~(\ref{Eq:W1}), we may infer that the
first-order Hamiltonian takes the form
$H^{(1)}_{{\gamma}J}=H_{\text{HFI}}$. Furthermore, from
Eq.~(\ref{Eq:W2_a}), we infer the second-order Hamiltonian to be
\begin{equation}
H^{(2)}_{{\gamma}J}=H_{\text{HFI}}R_{{\gamma}J}H_{\text{HFI}}
=\sum_{k_1,k_2}\left(T_{k_1}^{n}\cdot
T_{k_1}^{e}\right)R_{{\gamma}J}\left(T_{k_2}^{n}\cdot
T_{k_2}^{e}\right), \label{Eq:H2}
\end{equation}
where the latter expression is obtained by using
Eq.~(\ref{Eq:HFI_sp}) to represent $H_{\text{HFI}}$. The operator
$R_{{\gamma}J}$ here is the resolvent operator; it acts in the
electronic space and is given by the expressions
\begin{eqnarray*}
R_{{\gamma}J}&=&\sum_{{\gamma^\prime}{J^\prime}M_J^\prime}
\frac{\left\vert{\gamma^\prime}{J^\prime}{M_J^\prime}
\rangle\langle{{\gamma^\prime}{J^\prime}{M_J^\prime}}\right\vert}
{E_{{\gamma}J}-E_{{\gamma^\prime}{J^\prime}}}
\nonumber\\&=&\sum_{{\gamma^\prime}{J^\prime}{F^\prime}{M_F^\prime}}
\frac{\left\vert{\gamma^\prime}I{J^\prime}{F^\prime}{M_F^\prime}
\rangle\langle{{\gamma^\prime}I{J^\prime}{F^\prime}{M_F^\prime}}\right\vert}
{E_{{\gamma}J}-E_{{\gamma^\prime}{J^\prime}}}.
\end{eqnarray*}
It is important to realize that the resolvent operator behaves as a
scalar operator under rotations in the electronic space.

We may recouple the tensor operators of the second-order
Hamiltonian, Eq.~(\ref{Eq:H2}), to isolate parts acting in the
nuclear and electronic spaces. The resulting expression is
\begin{eqnarray*}
H^{(2)}_{{\gamma}J}&=&\sum_{k_1,k_2,k}\left(-1\right)^{k_1+k_2+k}\\&&\times
\left\{T_{k_1}^{n}\otimes{T_{k_2}^{n}}\right\}_k\cdot
\left\{T_{k_1}^{e}\otimes{R_{{\gamma}J}}T_{k_2}^{e}\right\}_k.
\end{eqnarray*}
(See the Appendix for notational descriptions; see
Ref.~\cite{VarMosKhe88} for general relations involving spherical
tensors.) The second-order correction to the energy is given by the
diagonal matrix elements of $H^{(2)}_{{\gamma}J}$ in the coupled
basis:
\begin{eqnarray}
W^{(2)}_F&=&
\langle{\gamma}IJFM_F\vert{H^{(2)}_{{\gamma}J}}\vert{\gamma}IJFM_F\rangle
\nonumber\\
&=&\left(-1\right)^{I+J+F}
\sum_{k_1,k_2,k}\left(-1\right)^{k_1+k_2+k}\left\{
\begin{array}{ccc}
I & J & F \\
J & I & k
\end{array}
\right\}\nonumber\\&&\times
\langle{I}\vert\vert{\left\{T_{k_1}^{n}\otimes{T_{k_2}^{n}}\right\}_k}
\vert\vert{I}\rangle \nonumber\\&&\times
\langle{\gamma}J\vert\vert
{\left\{T_{k_1}^{e}\otimes{R_{{\gamma}J}}T_{k_2}^{e}\right\}_k}
\vert\vert{\gamma}J\rangle. \label{Eq:W2}
\end{eqnarray}
Comparing this to Eq.~(\ref{Eq:W1}), we can see clearly that all
$k$-dependent terms enter the first- and second-order energy
expressions in an identical manner. However, because
$H_{\text{HFI}}$ does not contain a monopole contribution, the index
$k$ in Eq.~(\ref{Eq:W1}) is limited by $k>0$, whereas the summation
here is inclusive of the case $k=0$. Analyzing the $F$-dependent
factors entering this expression, namely the phase factor and
six-$j$ symbol, under the condition of $k=0$ gives
\begin{equation*}
\left(-1\right)^{I+J+F}\left\{
\begin{array}{ccc}
I & J & F \\
J & I & 0
\end{array}
\right\}=\frac{1}{\sqrt{(2I+1)(2J+1)}},
\end{equation*}
and we see that $F$-dependence is lost. Thus, the second-order $k=0$
terms here only cause an overall shift to the HFS energy levels, and
do not affect the interval spacing between levels.

If all second- and higher-order effects are negligible, then the HFS
intervals are determined completely by the terms $\langle
I||T_{k}^{n}||I\rangle\langle\gamma J||T_{k}^{e}||\gamma J\rangle$
appearing in Eq.~(\ref{Eq:W1}); we mention again that these terms
are related directly (see Eq.~(\ref{Eq:HFconsts})) to the
conventional hyperfine constants $A$ ($k=1$), $B$ ($k=2$), $C$
($k=3$), \dots. As $T_{k}^{n}$ and $T_{k}^{e}$ are spherical
tensors, the specific value of $k$ represents the underlying
rotational symmetry in the nuclear and electronic spaces,
respectively. Experimental measurements of the HFS interval spacings
may be used to determine these constants. However, if the
second-order effects cannot be neglected, then the terms $\langle
I||T_{k}^{n}||I\rangle\langle\gamma J||T_{k}^{e}||\gamma J\rangle$
can no longer be determined by the intervals. The ``constants''
which could be determined would also depend on second-order effects:
\begin{eqnarray*}
\lefteqn{\langle{I}\vert\vert{T_{k}^{n}}\vert\vert{I}\rangle
\langle{\gamma}J\vert\vert{T_{k}^{e}}\vert\vert{\gamma}J\rangle+
\sum_{k_1,k_2}\left(-1\right)^{k_1+k_2+k}}\\&&\times
\langle{I}\vert\vert{\left\{T_{k_1}^{n}\otimes{T_{k_2}^{n}}\right\}_k}
\vert\vert{I}\rangle \langle{\gamma}J\vert\vert
{\left\{T_{k_1}^{e}\otimes{R_{{\gamma}J}}T_{k_2}^{e}\right\}_k}
\vert\vert{\gamma}J\rangle.
\end{eqnarray*}
We note that second-order dipole-dipole ($k_1=k_2=1$) terms only
affect determination of the $A$ and $B$ constants, as two rank-one
tensors can only be coupled to form a tensor of rank $k\le2$. This
agrees with the proof given in Ref.~\cite{BelDerJoh08}, in which the
$C$ constant was shown not to be affected by the second-order
constant $\eta$. However, here we extend this conclusion to
dipole-dipole terms which mix states outside of the fine-structure
manifold as well. Similar conclusions can also be drawn here; for
example, second-order dipole-octupole terms do not affect the
determination of the $A$ constant, as a rank-one and a rank-three
tensor cannot be coupled to form a tensor of rank $k=1$.

The reduced matrix elements appearing here can be simplified, giving
\begin{eqnarray*}
\lefteqn{\langle{I}\vert\vert{\left\{T_{k_1}^{n}\otimes{T_{k_2}^{n}}\right\}_k}
\vert\vert{I}\rangle = \left(-1\right)^{2I+k}\sqrt{2k+1}}
\\&&\times\left\{
\begin{array}{ccc}
k_1 & k_2 & k \\
I   & I   & I
\end{array}
\right\} \langle{I}\vert\vert{T_{k_1}^{n}}\vert\vert{I}\rangle
\langle{I}\vert\vert{T_{k_2}^{n}}\vert\vert{I}\rangle,\\
\lefteqn{\langle{\gamma}J\vert\vert{\left\{T_{k_1}^{e}
\otimes{R_{{\gamma}J}}T_{k_2}^{e}\right\}_k}\vert\vert{\gamma}J\rangle
=\left(-1\right)^{2J+k}\sqrt{2k+1}}\\&&\times\sum_{{\gamma^\prime}J^\prime}
\left\{
\begin{array}{ccc}
k_1 & k_2 & k \\
J   & J   & J^\prime
\end{array}
\right\} \frac{
\langle{\gamma}J\vert\vert{T_{k_1}^{e}}\vert\vert{\gamma^\prime}J^\prime\rangle
\langle{\gamma^\prime}J^\prime\vert\vert{T_{k_2}^{e}}\vert\vert{\gamma}J\rangle}
{E_{{\gamma}J}-E_{{\gamma^\prime}{J^\prime}}}.
\end{eqnarray*}
These expressions may be used in Eq.~(\ref{Eq:W2}), at which point
the summation over $k$ may be carried out analytically by using a
well-known six-$j$ sum rule; the resulting expression is identical
to Eq.~(\ref{Eq:W2_a}).

The tensorial analysis of second-order HFI effects presented in this
Section could be applied to higher-order effects as well, with
similar insights following. For example, we could find that
experimental determination of the $D$ constant is not affected by
third-order dipole-dipole-dipole terms, due to the fact that three
rank-one tensors cannot be coupled to form a composite tensor with
the rotational symmetry of a rank-four tensor.

\section{Expressions for $P$-states of alkali metal atoms}
\label{Sec:HFSeqns} In this Section we concern ourselves with the
HFS equations of the lowest energy $P$-states of alkali-metal atoms,
namely the isotopes of Table \ref{Tbl:nucprops}. Specifically, we
are considering the $nP_{1/2}$ and $nP_{3/2}$ states, with
$n=2,3,4,5,6$ for isotopes of Li, Na, K, Rb, and Cs, respectively;
the specific $n$ will be implicit hereafter. As inferred from the
arguments of the previous Section, to determine the (first-order)
constants $A,B,C,\dots$ by measurement of HFS intervals, it is
necessary to have a knowledge of the higher-order effects. Here we
assume that these higher-order effects are sufficiently described by
the second-order dipole-dipole and dipole-quadrupole constants
$\eta$ and $\zeta$. These are defined in
Eqs.~(\ref{Eq:eta},\ref{Eq:zeta}); one convenience of these
definitions is that $\eta$ and $\zeta$ are identical for both the
$P_{1/2}$ and $P_{3/2}$ states (i.e.,
$\eta(P_{1/2})=\eta(P_{3/2})\equiv\eta$ and similar for $\zeta$).
Below we compile specific expressions for the first-order constants
in terms of the HFS intervals
$\delta{W}_F^{(J)}=W_F^{(J)}-W_{F-1}^{(J)}$ and the constants $\eta$
and $\zeta$.

For $^{6}$Li ($I=1$) we have
\begin{eqnarray*}
A(P_{1/2})&=& \frac{2}{3}\delta{W}_{3/2}^{(P_{1/2})}
             +\frac{1}{36}\eta+\frac{1}{12\sqrt{3}}\zeta, \\
A(P_{3/2})&=& \frac{1}{6}\delta{W}_{3/2}^{(P_{3/2})}
             +\frac{3}{10}\delta{W}_{5/2}^{(P_{3/2})}
             +\frac{1}{72}\eta-\frac{1}{120\sqrt{3}}\zeta, \\
B(P_{3/2})&=&-\frac{1}{3}\delta{W}_{3/2}^{(P_{3/2})}
             +\frac{1}{5}\delta{W}_{5/2}^{(P_{3/2})}
             +\frac{1}{36}\eta+\frac{1}{20\sqrt{3}}\zeta.
\end{eqnarray*}

For the isotopes $^{7}$Li, $^{23}$Na, $^{39}$K, $^{41}$K, and
$^{87}$Rb ($I=3/2$) we have
\begin{eqnarray*}
A(P_{1/2})&=& \frac{1}{2}\delta{W}_{2}^{(P_{1/2})}
             +\frac{1}{90}\eta+\frac{1}{15\sqrt{5}}\zeta, \\
A(P_{3/2})&=& \frac{1}{20}\delta{W}_{1}^{(P_{3/2})}
             +\frac{4}{25}\delta{W}_{2}^{(P_{3/2})}
             +\frac{21}{100}\delta{W}_{3}^{(P_{3/2})}
             \\&&+\frac{1}{180}\eta-\frac{1}{150\sqrt{5}}\zeta, \\
B(P_{3/2})&=&-\frac{1}{4}\delta{W}_{1}^{(P_{3/2})}
             -\frac{2}{5}\delta{W}_{2}^{(P_{3/2})}
             +\frac{7}{20}\delta{W}_{3}^{(P_{3/2})}
             \\&&+\frac{1}{30}\eta+\frac{1}{20\sqrt{5}}\zeta, \\
C(P_{3/2})&=&-\frac{1}{80}\delta{W}_{1}^{(P_{3/2})}
             -\frac{1}{100}\delta{W}_{2}^{(P_{3/2})}
             +\frac{1}{400}\delta{W}_{3}^{(P_{3/2})}
             \\&&+\frac{1}{400\sqrt{5}}\zeta.
\end{eqnarray*}

For $^{85}$Rb ($I=5/2$) we have
\begin{eqnarray*}
A(P_{1/2})&=& \frac{1}{3}\delta{W}_{3}^{(P_{1/2})}
             +\frac{1}{315}\eta+\frac{8}{105\sqrt{30}}\zeta, \\
A(P_{3/2})&=& \frac{3}{50}\delta{W}_{2}^{(P_{3/2})}
             +\frac{64}{525}\delta{W}_{3}^{(P_{3/2})}
             +\frac{9}{70}\delta{W}_{4}^{(P_{3/2})}
             \\&&+\frac{1}{630}\eta-\frac{4}{525\sqrt{30}}\zeta, \\
B(P_{3/2})&=&-\frac{1}{2}\delta{W}_{2}^{(P_{3/2})}
             -\frac{8}{21}\delta{W}_{3}^{(P_{3/2})}
             +\frac{15}{28}\delta{W}_{4}^{(P_{3/2})}
             \\&&+\frac{2}{63}\eta+\frac{1}{14\sqrt{30}}\zeta, \\
C(P_{3/2})&=& \frac{1}{40}\delta{W}_{2}^{(P_{3/2})}
             -\frac{1}{35}\delta{W}_{3}^{(P_{3/2})}
             +\frac{1}{112}\delta{W}_{4}^{(P_{3/2})}
             \\&&+\frac{3}{280\sqrt{30}}\zeta.
\end{eqnarray*}

Finally, for $^{133}$Cs ($I=7/2$) we have
\begin{eqnarray*}
A(P_{1/2})&=& \frac{1}{4}\delta{W}_{4}^{(P_{1/2})}
             +\frac{1}{756}\eta+\frac{1}{126}\zeta, \\
A(P_{3/2})&=& \frac{3}{56}\delta{W}_{3}^{(P_{3/2})}
             +\frac{2}{21}\delta{W}_{4}^{(P_{3/2})}
             +\frac{11}{120}\delta{W}_{5}^{(P_{3/2})}
             \\&&+\frac{1}{1512}\eta-\frac{1}{1260}\zeta, \\
B(P_{3/2})&=&-\frac{5}{8}\delta{W}_{3}^{(P_{3/2})}
             -\frac{1}{3}\delta{W}_{4}^{(P_{3/2})}
             +\frac{77}{120}\delta{W}_{5}^{(P_{3/2})}
             \\&&+\frac{1}{36}\eta+\frac{1}{120}\zeta, \\
C(P_{3/2})&=& \frac{1}{32}\delta{W}_{3}^{(P_{3/2})}
             -\frac{1}{24}\delta{W}_{4}^{(P_{3/2})}
             +\frac{7}{480}\delta{W}_{5}^{(P_{3/2})}
             \\&&+\frac{1}{480}\zeta.
\end{eqnarray*}

We note that in all of the above cases, the $\eta$ ($\zeta$)
contribution to $A(P_{3/2})$ is suppressed by a factor of $1/2$
($-1/10$) compared to its contribution to $A(P_{1/2})$.

\section{Results}\label{Sec:Results}
In order to calculate the second-order constants $\eta$ and $\zeta$
from Eqs.~(\ref{Eq:eta},\ref{Eq:zeta}), we must first generate the
$P_{1/2}$ and $P_{3/2}$ electronic states. This involves solving the
electron correlation problem. We employ an {\it ab initio}
relativistic coupled-cluster method which includes single, double,
and triple excitations from the lowest-order Dirac-Hartree-Fock
state. While both core and valence single and double excitations are
included, the triple excitations involve simultaneous excitation of
the valence electron with two core electrons. In other words, the
triple core excitations are not incorporated in the many-body
wavefunction. We refer to this method as the CCSDvT method. The most
sophisticated approximation within the CCSDvT method is described in
our previous work~\cite{DerPorBel08} for three-electron Li. For
heavier alkalis (beyond Li) the approximation of
Ref.~\cite{DerPorBel08} becomes computationally expensive and we
only keep the lowest-order terms on the right-hand-side of the
valence triples equation ($T_v[D_v]$ and $T_v[D_c]$ terms in
notation of Ref.~\cite{DerPorBel08}). For the electronic reduced
matrix elements entering Eqs.~(\ref{Eq:eta},\ref{Eq:zeta}), this
method gives results with accuracies better than 0.1\% for the
lightest (Li) and better than a few percent for the heaviest (Cs)
alkali-metal system.

Table~\ref{Tbl:mels} displays our computed values of the reduced
matrix elements of the electronic tensors $T_{1}^{e}$ and
$T_{2}^{e}$ between $P$ states, along with experimental values for
the fine structure energy splitting. The variation in these values
for different isotopes of the same atomic system is below the level
of precision shown. Combining these values with the nuclear data for
each isotope, Table~\ref{Tbl:nucprops}, we obtain the second-order
constants $\eta$ and $\zeta$. These are displayed in
Table~\ref{Tbl:corrections}. Also displayed in this Table are the
``corrections'' to the HFS constants $A(P_{1/2})$, $A(P_{3/2})$,
$B(P_{3/2})$, and $C(P_{3/2})$ deduced from $\eta$ and $\zeta$. Here
each correction is regarded as the difference between the actual
constant and the measured constant based only on first-order
equations. These can be inferred from the equations in
Section~\ref{Sec:HFSeqns}; e.g., for $^{133}$Cs:
$\Delta{A}(P_{1/2})=(1/756)\eta+(1/126)\zeta$.

\begin{table}[h]
\caption{Computed reduced matrix elements of the electronic tensors
$T_{1}^{e}$ and $T_{2}^{e}$ between $P$ states and experimental fine
structure intervals
$\Delta{E}_{\text{fs}}{\equiv}E_{P_{3/2}}-E_{P_{1/2}}$; fine
structure intervals are taken from
Ref.~\cite{Moo58}.}\label{Tbl:mels}
\begin{ruledtabular}\begin{tabular}{cddd}
Atom &
\multicolumn{1}{c}{$\langle{P_{3/2}}\vert\vert{T_{1}^{e}}\vert\vert{P_{1/2}}\rangle$}&
\multicolumn{1}{c}{$\langle{P_{3/2}}\vert\vert{T_{2}^{e}}\vert\vert{P_{1/2}}\rangle$}&
\multicolumn{1}{c}{$\Delta{E}_{\text{fs}}$} \\
& \multicolumn{1}{c}{(MHz/$\mu_N$)}& \multicolumn{1}{c}{(MHz/b)}&
\multicolumn{1}{c}{(cm$^{-1}$)} \\
\hline
Li &  19.0   &   11.9  &   0.3366  \\
Na &  12.1   &   59.4  &  17.1963  \\
K  & -12.6   & -103    &  57.600  \\
Rb &  20.1   &  213    & 237.598  \\
Cs &  29.2   &  313    & 554.11 \\
\end{tabular}\end{ruledtabular}
\end{table}

\begin{table*}[htb]
\caption{Second-order constants $\eta$ and $\zeta$ and corresponding
second-order corrections to the HFS constants. All values are in
kHz. $x[y]$ denotes $x\times10^y$.}\label{Tbl:corrections}
\begin{ruledtabular}\begin{tabular}{cm{2.5}m{2.5}m{2.5}m{2.5}m{2.5}m{2.5}m{2.5}}
Isotope & \multicolumn{1}{c}{$\eta$} & \multicolumn{1}{c}{$\zeta$} &
\multicolumn{1}{c}{$\Delta{A(P_{1/2})}$} &
\multicolumn{1}{c}{$\Delta{A(P_{3/2})}$} &
\multicolumn{1}{c}{$\Delta{B(P_{3/2})}$} &
\multicolumn{1}{c}{$\Delta{C(P_{3/2})}$} \\
\hline
$^{  6}$Li & 1.45[+2] & -2.02[-1] & 4.01     &  2.01     & 4.02     & \\
$^{  7}$Li & 2.53[+3] & -3.41[+1] & 2.70[+1] &  1.41[+1] & 8.34[+1] & -3.82[-2] \\
$^{ 23}$Na & 9.25     &  3.72     & 2.14[-1] &  4.03[-2] & 3.92[-1] &  4.16[-3] \\
$^{ 39}$K  & 9.33[-2] &  1.98[-1] & 6.94[-3] & -7.20[-5] & 7.54[-3] &  2.21[-4] \\
$^{ 41}$K  & 2.81[-2] &  1.32[-1] & 4.25[-3] & -2.38[-4] & 3.89[-3] &  1.48[-4] \\
$^{ 85}$Rb & 8.76[-1] &  2.68     & 4.01[-2] & -2.34[-3] & 6.28[-2] &  5.25[-3] \\
$^{ 87}$Rb & 2.87     &  2.57     & 1.08[-1] &  8.32[-3] & 1.53[-1] &  2.87[-3] \\
$^{133}$Cs & 3.51     & -6.69[-2] & 4.11[-3] &  2.38[-3] & 9.70[-2] & -1.39[-4] \\
\end{tabular}\end{ruledtabular}
\end{table*}

The most accurate reported value of the $^7$Li $A(P_{1/2})$ constant
based on experimental measurement of HFS intervals is that of Das
and Natarajan, Ref.~\cite{DasNat08}, wherein they give a value
$A(P_{1/2})=46.024(3)$ MHz. However, this value is based on
first-order HFS equations and is thus completely negligent of
higher-order effects. From Table~\ref{Tbl:corrections}, we see that
the second-order effects would cause a sizable shift of
$\Delta{A}(P_{1/2})=27.0$ kHz to this value. This shift is a full
order of magnitude larger than the claimed accuracy (3 kHz) of the
constant itself.
Das and Natarajan also claim accurate results for $A$ and $B$
constants for the $P$ states of all isotopes in Table
\ref{Tbl:nucprops}, with the exception of the $^{6,7}$Li $P_{3/2}$
and $^{41}$K $P_{1/2,3/2}$ states.

\begin{figure}[ht]
\begin{center}
\includegraphics[scale=0.45]{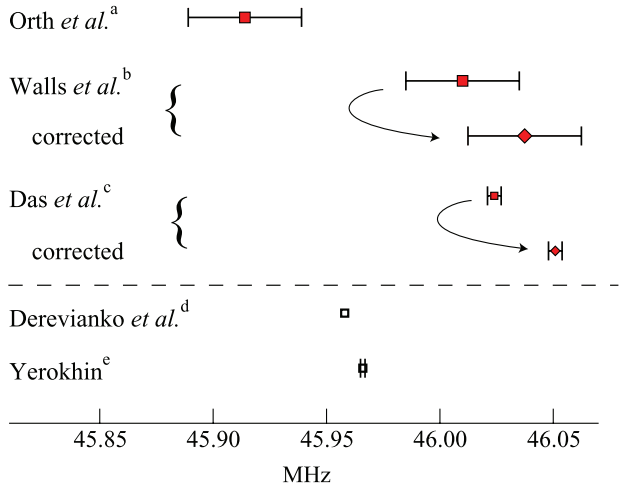}
\end{center}
\caption{(Color online) Values of the HFS constant $A(P_{1/2})$ for
$^7$Li. The heavy (hollow) squares denote experimental (theoretical)
values reported in the respective reference, and the heavy diamonds
denote the associated corrected values (of the immediately preceding
value) using second-order correction $\Delta{A}(P_{1/2})=27.0$ kHz
from Table~\ref{Tbl:corrections} when applicable. The uncertainty of
$\Delta{A}(P_{1/2})$ is negligible compared to the experimental
uncertainties in these cases. No uncertainty is given for $d$.
$^\text{a}$Ref.~\cite{OrtAckOtt75};%
$^\text{b}$Ref.~\cite{WalAshCla03};%
$^\text{c}$Ref.~\cite{DasNat08};%
$^\text{d}$Ref.~\cite{DerPorBel08};%
$^\text{e}$Ref.~\cite{Yer08}.%
\label{Fig:Aplot}}
\end{figure}

Figure~\ref{Fig:Aplot} displays various reported values of the
$^7$Li $A(P_{1/2})$ constant and their associated error bars. Along
with the value from Das and Natarajan, the Figure includes
experimental values from Orth {\it et al.}~\cite{OrtAckOtt75} and
Walls {\it et al.}~\cite{WalAshCla03}. The earlier value of Orth
{\it et al.}~accounted for second-order effects, whereas that of
Walls {\it et al.}~did not; both claim an accuracy of the same order
as our computed $\Delta{A}(P_{1/2})$. Also included in this Figure
are theoretical values recently reported by our
group~\cite{DerPorBel08} and Yerokhin~\cite{Yer08}. This Figure also
displays ``corrected'' values using our computed
$\Delta{A}(P_{1/2})$ for the applicable cases of Das and
Natarajan~\cite{DasNat08} and Walls {\it et al.}~\cite{WalAshCla03}.
We note that this correction causes these two values to shift
farther away from the values of Orth {\it et
al.}~\cite{OrtAckOtt75}, Derevianko {\it et al.}~\cite{DerPorBel08},
and Yerokhin~\cite{Yer08}; we do not attempt to explain the
resulting large discrepancy.

For the remaining HFS constants of Li, the second-order correction
again proves to be sizable, though not as pronounced as in the
$^7$Li $A(P_{1/2})$ case. For the $^6$Li $A(P_{1/2})$ constant, Das
and Natarajan claim an accuracy of the same order of magnitude as
our predicted second-order correction $\Delta{A}(P_{1/2})=4.01$ kHz.

Orth {\it et al.}~\cite{OrtAckOtt75,AriIngVio77} performed HFS
measurements of $^{6,7}$Li in the presence of strong fields and used
a complex fitting scheme to obtain the HFS constants. In addition to
the traditional constants, this scheme also yielded an off-diagonal
constant $A(P_{3/2},P_{1/2})$. This constant can be related to the
off-diagonal matrix element of the $T_{1}^{e}$ tensor by
\begin{equation*}
A(P_{3/2},P_{1/2})=\frac{1}{\sqrt{12}}
\left|\frac{\mu}{I}\langle{P_{3/2}||T^e_1||P_{1/2}}\rangle\right|.
\end{equation*}
This constant, along with the fine structure interval, parameterizes
the mixing between the two fine structure levels. In
Table~\ref{Tbl:Aod} we compare our value of this off-diagonal
constant for $^{7}$Li (deduced from the matrix element of
Table~\ref{Tbl:mels}) with the non-relativistic value from
Ref.~\cite{GuaWan98} and experimental value reported by Orth {\it et
al.}~\cite{OrtAckOtt75}. Our (relativistic) value is in close
agreement with both values.

\begin{table}[h]
\caption{Computed off-diagonal constant $A(P_{3/2},P_{1/2})$ for
$^{7}$Li compared to non-relativistic and experimental values.
Units are MHz.}\label{Tbl:Aod}
\begin{ruledtabular}\begin{tabular}{ld}
 &
\multicolumn{1}{c}{$A(P_{3/2},P_{1/2})$}\\
\hline
CCSDvT~{\it ab initio}; this work                       & 11.9      \\
non-relativ.~{\it ab initio}; Ref.~\cite{GuaWan98}  & 11.87853  \\
experiment; Ref.~\cite{OrtAckOtt75}                    & 11.823(81)
\end{tabular}\end{ruledtabular}
\end{table}

The second-order dipole-dipole correction was considered for
$^{133}$Cs in Ref.~\cite{GerDerTan03}, where a theoretical value of
$\eta=22.46$ kHz was used \footnote{The values of $\eta$ in this
paragraph were inferred from numbers within the cited reference, as
$\eta$ was not explicitly introduced in these papers.}. In a
following paper, Ref.~\cite{JohHoTan04}, this effect was reevaluated
using a more sophisticated third-order approach, resulting in a
value of $\eta=2.09$ kHz. Our present value ($\eta=3.51$ kHz), based
on the more complete CCSDvT method, is 68\% larger than the latter
value.

For the heavier alkali-metal atoms past Li, all of the predicted
second-order corrections (Table~\ref{Tbl:corrections}) are one to
six orders of magnitude smaller than the uncertainty in the most
precise reported experimental value of the respective HFS constant.
Thus, at the current levels of precision, the neglect of
second-order effects in these heavier systems is acceptable.
However, as experimental techniques improve and higher precision is
sought, it may become necessary to consider these higher-order
effects.

\section{Discussion}\label{Sec:Disc}
In this paper, we have demonstrated for the $P$ states of the
alkali-metal atoms that the second-order effects cannot simply be
neglected in deducing HFS constants from high-precision measurements
of the HFS intervals. This is exemplified by the $^7$Li $A(P_{1/2})$
constant, where the second-order effects are shown to cause a
$9\sigma$ shift to a recently reported value. Furthermore, we have
compiled values for the leading second-order HFS effects caused by
the dipole-dipole constants $\eta$ and the dipole-quadrupole
constants $\zeta$ for all naturally occurring alkali-metal isotopes.
These values can be used in conjunction with high-precision
measurements of the HFS intervals of the alkali-metal atoms to
determine the HFS constants.

This work was supported in part by the National Science Foundation.

\appendix

\section{Coupling of spherical tensors: notations and definitions}
\label{Sec:Append} Here we provide a description for the notations
used in Sections \ref{Sec:HFI-PT} and \ref{Sec:HFI-tens} along with
basic definitions involving coupling of spherical tensors. For a
more complete description of spherical tensors, including specific
formulae for recoupling of spherical tensors and matrix elements of
coupled tensor operators, see Ref.~\cite{VarMosKhe88}.

A spherical tensor of rank $k$, $P_k$, is a set of $2k+1$ operators.
We denote the individual components (operators) as $P_{kq}$, with
the index $q$ taking all integer values from $-k$ to $k$. Two
spherical tensors, $P_{k_1}$ and $Q_{k_2}$, may be coupled together
to form a composite spherical tensor of rank $k$. The rank of the
composite tensor is limited by
$\left|k_1-k_2\right|\leq{k}\leq{k_1+k_2}$. We denote the coupled
tensor by $\left\{P_{k_1}\otimes{Q_{k_2}}\right\}_k$; its components
are given by
\begin{equation}
\left\{P_{k_1}\otimes{Q_{k_2}}\right\}_{kq}=
\sum_{q_1q_2}C_{k_1q_1;k_2q_2}^{kq}P_{k_1q_1}Q_{k_2q_2}~,
\label{Eq:coupling}\end{equation}
where $C_{k_1q_1;k_2q_2}^{kq}$
represents Clebsch-Gordon coefficients. The scalar product of two
tensors of equal rank is defined as
\begin{eqnarray*}
P_{k}\cdot
Q_{k}&=&\left(-1\right)^k\sqrt{2k+1}\left\{P_{k}\otimes{Q_{k}}\right\}_{00}\\
&=&\sum_{q}\left(-1\right)^{q}P_{kq}Q_{k-q}~.
\end{eqnarray*}
If one of the tensors in Eq.~(\ref{Eq:coupling}) is a scalar, the
components of the coupled operator are simply
\begin{eqnarray*}
\left\{P_{0}\otimes{Q_{k}}\right\}_{kq}&=&P_{00}Q_{kq}~;\\
\left\{P_{k}\otimes{Q_{0}}\right\}_{kq}&=&P_{kq}Q_{00}~.
\end{eqnarray*}
When $P$ ($Q$) here is understood to be a scalar, the composite
tensor is then sufficiently represented by $PQ_k$ ($P_kQ$).

\bibliographystyle{apsrev}

\end{document}